\documentstyle[12pt]{article}
\begin{document}
\def\BbbZ{Z\!\!\!Z}
\begin{titlepage}
\begin{center}

\null
\vskip-1truecm
\rightline{IC/95/398}
\vskip1truecm
International Atomic Energy Agency\\
and\\
United Nations Educational Scientific and Cultural Organization\\
\medskip
INTERNATIONAL CENTRE FOR THEORETICAL PHYSICS\\
\vskip2truecm
{\bf OPERATOR REALIZATION OF THE $SU(2)$ WZNW MODEL\\}
\vskip2truecm
Paolo Furlan\\
Dipartimento di Fisica Teorica dell'Universit\`a di Trieste,
Trieste, Italy\\
and\\
Istituto Nazionale di Fisica Nucleare (INFN), Sezione di Trieste, Trieste,
Italy, \\
\bigskip
Ludmil K. Hadjiivanov\footnote{\normalsize On leave of absence from:
Institute for Nuclear Research and Nuclear Energy, Bulgarian Academy of
Sciences, Tsarigradsko Chaussee 72, BG--1784 Sofia, Bulgaria.}\\
International Centre for Theoretical Physics, Trieste, Italy\\
and\\
Istituto Nazionale di Fisica Nucleare (INFN), Sezione di Trieste, Trieste,
Italy,\\
\bigskip
and\\
\bigskip
Ivan T. Todorov$^1$\\
Department of Mathematics, MIT,
Cambridge, MA 02139--4307, USA.\\
\vskip2.5truecm
MIRAMARE -- TRIESTE\\
\medskip
December 1995\\
\end{center}

\end{titlepage}

\baselineskip=20pt

\centerline{ABSTRACT}
\bigskip

Decoupling the chiral dynamics in the canonical approach to the WZNW model
requires an extended phase space that includes left and right monodromy
variables $M$ and $\bar M$. Earlier work on the subject, which traced back the
quantum group symmetry of the model to the Lie--Poisson symmetry of the chiral
symplectic form, left some open questions:

-- How to reconcile the necessity to set $M\bar M^{-1}=1$ (in order to recover
the monodromy invariance of the local $2D$ group valued field $g=u\bar u$)
with
the fact the $M$ and $\bar M$ obey different exchange relations?

-- What is the status of the quantum symmetry in the $2D$ theory in which the
chiral fields $u(x-t)$ and $\bar u(x+t)$ commute?

-- Is there a consistent operator formalism in the chiral (and the extended
$2D$) theory in the continuum limit?

We propose a constructive affirmative answer to these questions for $G=SU(2)$
by presenting the quantum fields $u$ and $\bar u$ as sums of products of
chiral
vertex operators
and $q$--Bose creation and annihilation operators.

\newpage

\section{Introduction and summary of earlier results}

A key step in the treatment of the Wess--Zumino--Novikov--Witten (WZNW) model
[1], both axiomatic [2--4] and Lagrangean [5--9], is the construction of
chiral
vertex operators (CVO) and conformal current algebra blocks. While the
resulting 2-dimensional $(2D)$ braid invariant correlation functions satisfy
all Wightman axioms and can be used to reconstruct the quantum field operator
formalism in the physical state space, this is not the case for
the chiral theory. Trying to combine, following
[10, 11], the CVO of a given weight into a quantum
group tensor we have to abandon Hilbert space positivity. The alternative of
using (weak) quasi Hopf algebras [12] requires giving up coassociativity.

Here we continue our study [9] of the canonical approach to the problem (which
follows Gaw\c{e}dzki {\em et al.} [7]) specializing to the case $G=SU(2)$. We
proceed to summarizing relevant background and earlier results which will
serve
as a starting point for the present paper.

The general group valued periodic solution $g(t,x+2\pi )=g(t,x)$ of the WZNW
equations of motion factorizes into a product of right and left movers'
factors,
\setcounter{equation}{0}
\renewcommand{\theequation}{1.\arabic{equation}}
\begin{equation}
g(t,x)=u(x-t)\bar u(x+t),\quad u,\bar u\in SU(2) ,
\end{equation}
satisfying a weaker, {\em twisted periodicity} condition:
\begin{equation}
u(x+2\pi )=u(x)M,\quad \bar u(\bar x+2\pi )=\bar M^{-1}\ \bar
u(\bar x)\ .
\end{equation}
The symplectic form of the $2D$ theory can be presented as a sum of two
decoupled closed chiral 2--forms at the price of considering the monodromy
matrices $M$ and $\bar M$ as independent of each other additional dynamical
variables. One then derives [7] quadratic Poisson bracket relations for the
components $M_\pm$ of the Gauss decomposition of $M$:
\begin{equation}
M=M_+M^{-1}_-,\quad M_+=e^{-i\pi\Delta}\left(\matrix{ d &b\cr \cr
0&d^{-1}\cr}\right) ,\quad
M_-=e^{i\pi\Delta}\left(\matrix{ d^{-1} &0\cr \cr
c&d\cr}\right)
\end{equation}
($\Delta ={3\over 4h}$ being the conformal dimension of $u$,
see Eq.(1.13)) -- and similarly
for the bar variables. Using the tensor product notation of Faddeev {\em et
al.} [13], $\stackrel{1}{u} =u\otimes 1\!\!1$, $ \stackrel{2}{u} =
1\!\!1\otimes u$, one can write the quantized exchange relations in the
form [7, 9]
\begin{equation}
\stackrel{2}{u}(x_2)\stackrel{1}{u}(x_1)=\stackrel{1}{u}(x_1)\stackrel{2}{u}
(x_2)R(x_{12}),\quad
x_{12}=x_1-x_2
\end{equation}
where the quantum $R$--matrix is given by
\begin{equation}
R(x)=R^-\ \theta (x)+R^+\ \theta (-x)\ .
\end{equation}
The step function $\theta$ on the infinite cover of the circle is assumed to
have a periodic derivative:
\begin{equation}
2\pi\theta '(x)=\delta (x)=
\sum^\infty_{n=-\infty}\ e^{inx},\quad
\theta (x)+\theta (-x)=1 \ .
\end{equation}
$R^\pm$ are $4\times 4$ matrices
solving (properly extended on the tensor cube of spaces)
the Yang--Baxter equation
\begin{equation}
R^\varepsilon_{12}\
R^\pm_{13}\ R^\pm_{23}=R^\pm_{23}\ R^\pm_{13}\ R^\varepsilon_{12},\quad
\varepsilon =+ , - \ .
\end{equation}
They yield (upon multiplication with a permutation $P\ ,\
P (x\otimes y)=y\otimes x$) a pair of braid
operators with inverse eigenvalues:
\begin{equation}
\check R^\pm=R^\pm P=q^{\mp{1\over 2}}\Pi_3-q^{\pm{3\over
2}}\Pi_1\ .
\end{equation}
Here $\Pi_i\ \  (i=1,3)$ are $i$--dimensional orthogonal projectors,
\begin{equation}
\Pi^2_i=\Pi_i=\Pi^*_i\ ,\ \ \  \Pi_1\Pi_3=0\ ,\ \ \
\Pi_1+\Pi_3=1\ ,\
\ \ {\rm tr}\ \Pi_i=i\ (=1,3)\ .
\end{equation}
If we introduce the $SL_q(2)$ invariant ``$q$--skew symmetric'' tensor
\setcounter{equation}{0}
\renewcommand{\theequation}{1.10\alph{equation}}
\begin{equation}
\left({\cal E}_{\alpha\beta}\right) =\left(\matrix{ 0 &-q^{1/2}\cr \cr
\bar q^{1/2} &0\cr}\right)\ \ ,\quad \bar q =q^{-1}
\end{equation}
and its inverse
\begin{equation}
\left({\cal E}^{\alpha\beta}\right) =\left( -{\cal
E}_{\alpha\beta}\right),\quad {\cal E}^{\alpha\sigma}{\cal
E}_{\sigma\beta}=\delta^\alpha_\beta\ ,
\end{equation}
then we can write
\setcounter{equation}{0}
\renewcommand{\theequation}{1.11\alph{equation}}
\begin{eqnarray}
{{\Pi}_1}^{\alpha\beta}_{\rho\sigma} &=&-{1\over [2]}\ {\cal E}^{\alpha\beta}
{\cal E}_{\rho\sigma}\ \ \ \  \left( -{\cal E}^{\alpha\beta}{\cal
E}_{\alpha\beta}=[2]:=q+\bar q\right)\\
{{\Pi}_3}^{\alpha\beta}_{\rho\sigma} &=&\delta^\alpha_\rho\delta^\beta_\sigma
+{1\over [2]}\ {\cal E}^{\alpha\beta}{\cal E}_{\rho\sigma}\quad.
\end{eqnarray}
The condition that the eigenvalues of $\check R^-$ coincide with those of the
braid matrices (computed from the conformal and $SU(2)$ invariant 3--point
functions) fixes the value of $q$; in particular, the triply degenerate
eigenvalue is
\setcounter{equation}{11}
\renewcommand{\theequation}{1.\arabic{equation}}
\begin{equation}
q^{1/2}=\exp\left\{ i\pi\left(\Delta_1-2\Delta_{1/2}\right)\right\}
=\exp\left({\pi i\over 2h}\right)\ ,\ \ \ h=k+2
\end{equation}
where $k$ is the Kac-Moody level,
the conformal dimension $\Delta_I$ for a CVO of isospin $I$
(at height $h$) being
\begin{equation}
\Delta_I={1\over h}\ I(I+1)\ \ .
\end{equation}
The $R$--matrix so obtained yields the correct $k\to\infty$ limit in terms of
classical Poisson brackets.

The exchange relations involving the triangular factors $M_\pm$ of the
monodromy matrix (1.3) are also expressed in terms of $R^\pm$:
\begin{equation}
\stackrel{1}{M}_\pm\stackrel{2}{u}(x)=\ \stackrel{2}{u}(x)R^\pm
\stackrel{1}{M}_\pm\ ,
\end{equation}
\setcounter{equation}{0}
\renewcommand{\theequation}{1.15\alph{equation}}
\begin{eqnarray}
R^\varepsilon\stackrel{1}{M}_\pm\stackrel{2}{M}_\pm  &=&
\stackrel{2}{M}_\pm\stackrel{1}{M}_\pm R^\varepsilon\ ,\quad \varepsilon
=+ , - \ ,\\
R^\pm\stackrel{1}{M}_\pm\stackrel{2}{M}_\mp &=&
\stackrel{2}{M}_\mp\stackrel{1}{M}_\pm R^\pm \ .
\end{eqnarray}

The symmetry of (1.4) and (1.14) under (local) left shifts of $u(x)$ is
generated by a periodic chiral current
$j(x)=j^a(x)\sigma_a\in su_2$ such that
\setcounter{equation}{15}
\renewcommand{\theequation}{1.\arabic{equation}}
\begin{eqnarray}
{[}\stackrel{1}{u}(x_1),\stackrel{2}{j}(x_2)] &=& C\stackrel{1}{u}(x_1)
\delta (x_{12}) ,\\
{[}\stackrel{1}{j}(x_1),\stackrel{2}{j}(x_2)] &=& [C , \stackrel{1}{j}(x_1)]
\delta (x_{12})+ik C \delta '(x_{12})\ .
\end{eqnarray}
Here $C$ is the Casimir invariant
\begin{equation}
C={1\over 2}\
\stackrel{1}{\sigma} _a\stackrel{2}{\sigma} _a=P-{1\over 2}\ 1\ .
\end{equation}
Classically the current $j$ is expressed in terms of the group valued field
$u$
as $j=-ik\ u'u^{-1}$. The quantum version of this relation is the operator
Knizhnik--Zamolodchikov equation [2, 3]:
\begin{equation}
-ihu'(x)=\ :j(x)u(x):\ ,
\end{equation}
where the normal product in the right--hand side is defined in terms of the
frequency parts of $j$:
$$:j(x)u(x):\ =\ \sigma_a\left\{ j^a_{(+)}(x)u(x)+u(x)
j^a_{(-)}(x)\right\} \ ,\nonumber $$
\begin{equation}
j^a_{(-)}(x)=\sum^\infty_{n=0} J^a_ne^{inx}\ ,\quad
j^a_{(+)}(x)=\sum^{-1}_{n=-\infty} J^a_ne^{inx}\ ,
\end{equation}
$$j(x) = j_{(+)}(x)+j_{(-)}(x)\ ,\ \  j_{(-)}(x)\vert 0\rangle =0=\langle
0\vert j_{(+)}(x) \ .\nonumber $$
Similar, although not identical, relations are derived [7, 9]
for the left mover
(bar) sector:
\setcounter{equation}{0}
\renewcommand{\theequation}{1.21\alph{equation}}
\begin{equation}
\stackrel{2}{\bar u}(\bar x_2)\stackrel{1}{\bar u}(\bar x_1) =
{\bar R}(\bar x_{21}) \stackrel{1}{\bar u}(\bar x_1)\stackrel{2}{\bar
u}(\bar x_2) \ ,\\
\end{equation}
where
\begin{equation}
{\bar R}({\bar x}) = {\bar R}^-\  \theta ({\bar x}) + {\bar R}^+\
\theta (-{\bar x})
\end{equation}
and ${\bar R}^{\pm}$ are related to $R^{\pm}$ by
\begin{equation}
{\bar R}^{\pm} = PR^{\pm} P\ ,\ \ {\bar R}^{\pm} R^{\mp} = 1\ ;
\end{equation}
\setcounter{equation}{21}
\renewcommand{\theequation}{1.\arabic{equation}}
\begin{equation}
\stackrel{1}{\bar M}_\pm\stackrel{2}{\bar u}(\bar x) =R^\pm
\stackrel{2}{\bar u}(\bar x)\stackrel{1}{\bar M}_\pm \ ;\\
\end{equation}
\setcounter{equation}{0}
\renewcommand{\theequation}{1.23\alph{equation}}
\begin{equation}
\stackrel{1}{\bar M}_\pm\stackrel{2}{\bar M}_\pm R^\varepsilon =
R^\varepsilon\stackrel{2}{\bar M}_\pm\stackrel{1}{\bar M}_\pm\ ,
\ \ \varepsilon = + , -\\
\end{equation}
\begin{equation}
\stackrel{1}{\bar M}_\pm\stackrel{2}{\bar M}_\mp R^\pm =
R^\pm\stackrel{2}{\bar M}_\mp\stackrel{1}{\bar M}_\pm\ ;
\end{equation}
\setcounter{equation}{23}
\renewcommand{\theequation}{1.\arabic{equation}}
\begin{equation}
{[}\stackrel{1}{\bar j}(\bar x_1), \stackrel{2}{\bar u} (\bar x_2)]=\delta
(\bar x_{12})
\stackrel{2}{\bar u} (\bar x_2)C\ .
\end{equation}
The left and right sectors are completely decoupled, their dynamical variables
commute between each other:
\begin{equation}
{[}\stackrel{1}{u}(x), \stackrel{2}{\bar u}(\bar
x')]=0=[\stackrel{1}{M}_\varepsilon ,\stackrel{2}{\bar M}_{\varepsilon '}]=
{[}\stackrel{1}{u}(x),\stackrel{2}{\bar M}_\varepsilon
{]}=[\stackrel{1}{M}_\varepsilon ,\stackrel{2}{\bar u}(\bar
x)]\ .
\end{equation}
It is instructive to verify that the above exchange relations imply the local
commutativity of $g$ (1.1) :
\begin{equation}
{[}{\stackrel{1}g}(t_1,x_1),{\stackrel{2}g}(t_2,x_2)]=0\ \
{\rm for}\ \  (x_{12}-t_{12})(x_{12}+t_{12})>0\ .
\end{equation}

There appears to be a price for the decoupling of the left and right dynamics.
There is a difference between the monodromy exchange relations (1.15) and
(1.23) (which can be traced back to a sign difference between the
corresponding
classical Poisson brackets -- see [9]). Hence, {\em one cannot identify the
dynamical variables $M$ and $\bar M$}. The question arises: {\em can we then
recover the monodromy invariance (i.e., the single valuedness) of the $2D$
field (1.1)} ? A related problem emerges in trying to make precise the quantum
group invariance of $g$ (see Sec.2 below). The clue to the solution of these
problems lies in the realization that the left--right extended WZNW model has
features of a gauge quantum field theory. The physical Hilbert space of the
$2D$ theory is a proper subquotient
of the tensor product of state spaces of the
chiral theories. The monodromy invariance of the product (1.1) is recovered in
a weak sense -- as an equation for matrix elements between physical states.
The
quantum group properties of the $2D$ theory are examined in similar terms. We
start by factorizing the dependence of the chiral field $u(x)=
( u^\alpha_\beta (x) )$
on its quantum group index $\beta$ and use the ensuing $q$--Bose creation and
annihilation operators to express the monodromy matrices $M_\pm$.

\setcounter{subsection}{0}
\renewcommand{\thesubsection}{2\Alph{subsection}}
\section{Operator realization of the chiral exchange relations}
\subsection{Quantum group symmetry. Basic building blocks: the $U_q$
oscillators}

The Lie--Poisson symmetry of the quadratic Poisson brackets of $u(x)$ and
$M_\pm$ [6, 7] gives rise to a ``quantum symmetry'' under $GL_q(2)$
transformation
\setcounter{equation}{0}
\renewcommand{\theequation}{2.\arabic{equation}}
\begin{equation}
u(x)\to u(x)T,\quad M\to T^{-1}MT
\end{equation}
where the $2\times 2$ matrix $T$ has non--commuting elements characterized by
the exchange relations [13]
\setcounter{equation}{0}
\renewcommand{\theequation}{2.2\alph{equation}}
\begin{equation}
\stackrel{1}{T} \stackrel{2}{T} R^\varepsilon =
R^\varepsilon \stackrel{2}{T}
\stackrel{1}{T} ,\quad \varepsilon =+,-
\end{equation}
or
\begin{equation}
T^\beta_2T^\beta_1=qT^\beta_1T^\beta_2\ ,\  T^2_\beta T^1_\beta =qT^1_\beta
T^2_\beta\ ,\ [T^1_2,T^2_1]=0\ ,\ [T^1_1,T^2_2]=(\bar
q-q)T^1_2T^2_1\ .
\end{equation}
$GL_q(2)$ has for generic $q$ a 1--dimensional centre generated by the
$q$--determinant:
\setcounter{equation}{0}
\renewcommand{\theequation}{2.3\alph{equation}}
\begin{equation}
{\rm det}_qT=T^1_1T^2_2-\bar qT^1_2T^2_1=T^2_2T^1_1-qT^1_2T^2_1\ ;
\end{equation}
this allows to define the factor algebra
\begin{equation}
SL_q(2)=\{ T\in GL_q(2);\  {\rm det}_qT=1\}\ .
\end{equation}
It follows from (2.2) that the tensor product of quantum group matrices
commutes with the braid operators (1.8):
\setcounter{equation}{3}
\renewcommand{\theequation}{2.\arabic{equation}}
\begin{equation}
\left[\check R^\pm\ ,\stackrel{1}{T}\stackrel{2}{T}\right] =0\
\Leftrightarrow
\ \left[\Pi_i\ ,\stackrel{1}{T}\stackrel{2}{T}\right] =0\ ,\ \ \
i=1,3\ .
\end{equation}
The entries of $T$ can be viewed as linear functionals on the quantum
universal
enveloping algebra (QUEA) $U_q=U_q(s\ell_2)$ [13].
\vskip 1cm

An elementary realization of the exchange relations (1.4) is given by two
(conjugate) pairs of
$SL_q(2)$ covariant oscillators
$a^\pm_\alpha$ satisfying
\setcounter{equation}{0}
\renewcommand{\theequation}{2.5\alph{equation}}
\begin{eqnarray}
a^\varepsilon_\alpha a^\varepsilon_\beta &=& q^{\pm{1\over 2}}\
a^\varepsilon_\rho a^\varepsilon_\sigma\ (\check R^{\pm} )^{\rho\sigma}
_{\alpha\beta}\\
a^-_\alpha a^+_\beta &=& q^{{1\over 2}} a^+_\rho a^-_\sigma\ (\check
R^+ )^{\rho\sigma}_{\alpha\beta} +\bar q^N{\cal E}_{\alpha\beta}\ ,
\end{eqnarray}
where ${\cal E}_{\alpha\beta}$ is the $SL_q(2)$ invariant tensor (1.10),
\setcounter{equation}{5}
\renewcommand{\theequation}{2.\arabic{equation}}
\begin{equation}
{\cal E}_{\rho\sigma}T^\rho _\alpha T^\sigma _\beta ={\cal
E}_{\alpha\beta}\left( =-q^{\mp{3\over 2}}\ {\cal E}_{\rho\sigma}(\check
R^{\pm} )^{\rho\sigma}_{\alpha\beta}\right)\ .
\end{equation}
We shall interpret $a^+_\alpha$ and $a^-_\alpha$ as creation and annihilation
operators setting
\begin{equation}
a^-_\alpha\vert 0\rangle =0=\langle 0\vert a^+_\beta\ \ \Rightarrow
\ \ a^-_\alpha
a^+_\beta\vert 0\rangle ={\cal E}_{\alpha\beta}\vert
0\rangle\ .
\end{equation}
The
$SL_q(2)$ invariant
combinations of $a^\pm_\alpha$ are expressed in terms of
the {\em number operator} $N$ determined (mod $2h$ for $q^h=-1$) by
\begin{equation}
q^N\ a^\pm_\alpha =a^\pm_\alpha\ q^{N\pm 1}\ ,\quad (q^N-1)\vert
0\rangle =0\ ;
\end{equation}
we have
\setcounter{equation}{0}
\renewcommand{\theequation}{2.9\alph{equation}}
\begin{equation}
a^+_\alpha{\cal E}^{\alpha\beta}a^-_\beta =[N]:={q^N-\bar
q^N\over q-\bar q}\ ,\ \
a^-_\alpha{\cal E}^{\alpha\beta} a^+_\beta =-[N+2]\ ,
\end{equation}
\begin{equation}
a^\pm_\alpha{\cal E}^{\alpha\beta}a^\pm_\beta =0\ \
\Leftrightarrow \ \ a^{\pm}_2a^{\pm}_1=qa^{\pm}_1a^{\pm}_2\ .
\end{equation}
The $SL_q (2)$ invariance of the exchange relations (2.5) is
equivalent to their $U_q$ invariance [14]. Introducing the raising
and lowering Chevalley generators $E$ and $F$ such that
\setcounter{equation}{0}
\renewcommand{\theequation}{2.10\alph{equation}}
\begin{equation}
[E, F] = [H]\ ,\ \ q^H E = E q^{H+2}\ ,\ \ q^H F = F q^{H-2}
\end{equation}
and defining their coproduct by
\begin{equation}
\Delta (E) = E\otimes q^H + 1\otimes E\ , \ \
\Delta (F) = F\otimes 1 + {\bar q}^H\otimes F\ ,\ \
\Delta (q^{\pm H}) = q^{\pm H}\otimes q^{\pm H}\ ,
\end{equation}
we verify that the relations (2.5) are invariant under the
following $U_q$ transformation law:
\setcounter{equation}{0}
\renewcommand{\theequation}{2.11\alph{equation}}
\begin{eqnarray}
&&q^H a^{\pm}_1 = a^{\pm}_1 q^{H+1}\ \ ,\ \ q^H a^{\pm}_2 =
a^{\pm}_2 q^{H-1}\\
&&[ E, a^{\pm}_1 ] = 0 = F a^{\pm}_2 - q a^{\pm}_2 F\ \ ,\ \
[ E, a^{\pm}_2 ] = a^{\pm}_1 q^H \ \ , \ \
F a^{\pm}_1 - {\bar q} a^{\pm}_1 F = a^{\pm}_2 \ .
\end{eqnarray}
The $U_q(gl_2)$ Cartan subalgebra generated by $q^H$ and $q^N$ involves
the individual number operators
$N_\alpha \ ,\
\alpha =1,2\ ,$
satisfying
\setcounter{equation}{0}
\renewcommand{\theequation}{2.12\alph{equation}}
\begin{equation}
\  N_1+N_2=N\ ,\ \ \ N_1-N_2=H
\end{equation}
and consequently, in view of (2.8), (2.11a),
\begin{equation}
q^{N_\alpha}\ a^+_\beta =a^+_\beta\ q^{N_\alpha +\delta_{\alpha\beta}}\ ,\ \
[N_{\alpha} ,a_1^+ a_2^- ]=0=[N_{\alpha} ,a_2^+ a_1^- ]\ \ .
\end{equation}
The Fock space of the q-oscillator algebra (with an $U_q$
invariant vacuum vector satisfying (2.7)) possesses two $U_q$
invariant forms: a hermitean (sesquilinear) one, $(~,~)$ , and a
bilinear one, $\langle ~,~\rangle$.
There are, accordingly, two antiinvolutions
(i.e. involutive algebra antihomomorphisms) that extend the known
ones for $U_q$ -- cf. [15].
One can define -- for any $O$ in the oscillator algebra --
an {\em antilinear hermitean conjugation} $O\to O^*$ for which
\setcounter{equation}{0}
\renewcommand{\theequation}{2.13\alph{equation}}
\begin{equation}
E^*=F\ ,\ \ F^*=E\ ,\ \ (q^H)^*=\bar q^H
\end{equation}
(and $\Delta (X^*)=\Delta (X)^*$ for $(X_1\otimes X_2)^*
=X^*_2\otimes X^*_1\ ,\ \forall X, X_1 ,X_2 \in U_q)\ $
such that the following
counterpart of the familiar relation between (undeformed)
creation and annihilation, and number operators holds:
\begin{equation}
a^+_\alpha (a^+_\alpha )^*=[N_\alpha ]\ ,\ \ \ (a^+_\alpha )^*a^+_\alpha
=[N_\alpha +1]\ ,\ \ \ \alpha =1,2\ ,
\end{equation}
and a {\em linear transposition} $O\to\ ^tO$, satisfying
\setcounter{equation}{0}
\renewcommand{\theequation}{2.14\alph{equation}}
\begin{equation}
^tE = Fq^H\ ,\ \  ^tF=\bar q^HE\ , \ \  ^t(q^H)=q^H
\end{equation}
(so that $\Delta ({^tX})={ ^t\Delta}(X)$ for
$^t(X_1 \otimes X_2 )= {^tX}_1\otimes {^tX}_2\ $),  and
\begin{equation}
\sum^2_{\alpha =1}\ a^+_\alpha\ ^t(a^+_\alpha )=[N]\ ,\ \
\sum^2_{\alpha =1}\ a^-_\alpha\ ^t(a^+_\alpha )=0\ \ .
\end{equation}
It is an easy exercise to verify that (2.13b) is satisfied if we set
\setcounter{equation}{14}
\renewcommand{\theequation}{2.\arabic{equation}}
\begin{equation}
(a^+_1)^*=q^{{1\over 2}+N_2}a^-_2=:a_1\ ,\ \
(a^+_2)^*=-\bar q^{{1\over 2}+N_1}a^-_1=:a_2\ ,\ \
a^*_{\alpha}=a^+_{\alpha}\ ,\ \
(q^{N_\alpha})^*=\bar q^{N_\alpha}\ ,
\end{equation}
while (2.14b) will take place if
\setcounter{equation}{0}
\renewcommand{\theequation}{2.16\alph{equation}}
\begin{equation}
^t(a^+_\alpha )={\cal E}^{\alpha\beta}a^-_\beta =:a^{\alpha} ,\ \ \
^t(a^-_\alpha )=-{\cal E}^{\alpha\beta}a^+_\beta\ ,\ \ \
^t(q^{N_{\alpha}})=q^{N_{\alpha}}\ ,
\end{equation}
so that
\begin{equation}
a^1={\bar q}^{N_2}a_1\ ,\ \ a^2=q^{N_1}a_2\ .
\end{equation}
The (infinite dimensional)
Fock space $\widetilde{\cal F}$ is spanned by the vectors
\setcounter{equation}{0}
\renewcommand{\theequation}{2.17\alph{equation}}
\begin{equation}
\Phi_{n_1n_2}=(a^+_1)^{n_1}(a^+_2)^{n_2}\vert 0\rangle\ ,
\quad n_1,n_2\in\BbbZ
\end{equation}
that form an orthogonal basis with respect to both forms:
\begin{eqnarray}
(\Phi_{n_1n_2},\Phi_{m_1m_2}) &=& \langle 0\vert
a^{n_2}_2a^{n_1}_1(a^*_1)^{m_1}(a^*_2)^{m_2}\vert 0\rangle =\nonumber\\
&=&\delta_{n_1m_1}\delta_{n_2m_2}[n_1]![n_2]!\\
\langle\Phi_{n_1n_2},\Phi_{m_1m_2}\rangle
&=& \langle 0\vert
(a^2)^{n_2}(a^1)^{n_1}(a^+_1)^{m_1}(a^+_2)^{m_2}\vert 0\rangle
=\nonumber\\
&=&\delta_{n_1m_1}\delta_{n_2m_2}\bar
q^{n_1 n_2}[n_1]![n_2]!\ \ .
\end{eqnarray}
The sesquilinear form (2.17b) is real but not positive definite (for $q\bar
q=1, q\not= 1$). For $q^{1/2}$ given by (1.12) $(q^h=-1)$ it is {\em
semidefinite}: $\widetilde{\cal F}$
contains an infinite dimensional subspace $\widetilde{\cal
F}^{(0)}$ of {\em null (zero norm) vectors} spanned by $\vert
n_1,n_2\rangle$ with $max (n_1,n_2)\geq h$. Setting
\setcounter{equation}{17}
\renewcommand{\theequation}{2.\arabic{equation}}
\begin{equation}
(a^\pm_\alpha )^h=0 \ ,
\end{equation}
one obtains a {\em finite} $( h^2 )$ dimensional Fock quotient space
${\cal F}_h = {\widetilde {\cal F}}/{{\widetilde {\cal F}}^{(0)}}$
on which the sesquilinear form (2.17b) is already positive definite. ${\cal
F}_h$ splits into a direct sum of unitary irreducible representations of
$U_q$ with spins $0\leq I\leq {h-1\over 2}$. Except for
$I={h-1\over 2}$, all other spins appear twice -- in a ``standard'' and
a ``shadow'' representation differing by the eigenvalue of $q^N$.

One can identify the $U_q$
generators $E$, $F$ and $q^{\pm H}$with
\begin{equation}
E=a^*_1a_2\ q^{N_1}\ ,\quad F=\bar q^{N_1}\ a^*_2a_1\ ,\quad
q^{\pm H}
=q^{\pm (N_1-N_2 )}\ .
\end{equation}
In verifying the consistency between (2.19) and (2.10a), (2.11) one uses
(2.12b), (2.13b) and the definitions (2.5), (2.15) which imply
\begin{equation}
a_2a_1=qa_1a_2\ ,\ \ \  a_1a^*_2=qa^*_2a_1\ \  (\ \Rightarrow
\ a^*_2a^*_1=qa^*_1a^*_2\ ,\ \ \
a_2a^*_1=\bar qa^*_1a_2\ )\ .
\end{equation}
We note that (2.10a), (2.11) and (2.19)
(implying
$EF=[N_1]\ [N_2+1]$, $FE=[N_1+1]\
[N_2]\ $) yield the basic (anti) commutation relations of the Chevalley
generators of the quantum superalgebra $U_{q^{1/2}}(osp(1,4))$.
We are, in fact, dealing with its $(h^2$ dimensional Fock space) {\em
singleton
representation} [16].

\subsection{Factorized form of $u(x)$. Monodromy matrices}

The correlation functions of the chiral field $u(x)=(u^\alpha_\beta (x))$ can
be reconstructed if we factorize the dependence on the $SU(2)$ index $\alpha$
and the quantum group index $\beta$ setting
\setcounter{equation}{0}
\renewcommand{\theequation}{2.21\alph{equation}}
\begin{equation}
u^\alpha_\beta (x)=u^\alpha_+(x,N)a^+_\beta +a^-_\beta\ u^\alpha_-(x,N) ,
\end{equation}
with
\begin{equation}
u_-\vert 0\rangle =0=\langle 0\vert u_+\ \ .
\end{equation}
It is assumed that $u_\pm$ and $a^\pm$ are only coupled through the number
operator:
\setcounter{equation}{21}
\renewcommand{\theequation}{2.\arabic{equation}}
\begin{equation}
a^\varepsilon_\beta\ u^\alpha_{\varepsilon '} (x,N)=u^\alpha_{\varepsilon
'}(x,N-\varepsilon )a^\varepsilon_\beta\ \ .
\end{equation}
Inserting (2.21) into (1.4) and using (2.5) we ``diagonalize'' the exchange
relations for the CVO $u^\alpha_\varepsilon$. The result is particularly
simple
for an equal frequency pair:
\setcounter{equation}{0}
\renewcommand{\theequation}{2.23\alph{equation}}
\begin{equation}
u^{\alpha_2}_\pm(x_2,N\pm 1)u^{\alpha_1}_\pm (x_1,N)=q^{{1\over 2}\varepsilon
(x_{12})}\ u^{\alpha_1}_\pm (x_1,N\pm 1)u^{\alpha_2}_\pm (x_2,N)
\end{equation}
where
\begin{equation}
\varepsilon (x) =\theta (x)-\theta (-x)\ \ .
\end{equation}
For a product of opposite frequency $u$'s we introduce a (symmetric) 3--vector
and a (skewsymmetric) scalar bilocal combination (obtained by applying the
projectors $\Pi_3$ and $\Pi_1$ (1.11), respectively):
\setcounter{equation}{0}
\renewcommand{\theequation}{2.24\alph{equation}}
\begin{eqnarray}
V^{\alpha_1\alpha_2}(x_1,x_2;N) &=& u^{\alpha_1}_+(x_1,N)\
u^{\alpha_2}_-(x_2,N)+\nonumber \\
&&+qu^{\alpha_1}_-(x_1,N+1)u^{\alpha_2}_+(x_2,N+1)\\
S^{\alpha_1\alpha_2}(x_1,x_2;N) &=& u^{\alpha_1}_+(x_1,N)\
u^{\alpha_2}_-(x_2,N)[N]-\nonumber\\
&&-u^{\alpha_1}_-(x_1,N+1)u^{\alpha_2}_+(x_2,N+1)[N+2]\ \ .
\end{eqnarray}
These are again just multiplied by a phase, under an exchange of the
arguments:
\setcounter{equation}{0}
\renewcommand{\theequation}{2.25\alph{equation}}
\begin{eqnarray}
V^{\alpha_2\alpha_1}(x_2,x_1;N) &=& q^{{1\over 2}\varepsilon (x_{12})}\
V^{\alpha_1\alpha_2}(x_1,x_2;N)\\
S^{\alpha_2\alpha_1}(x_2,x_1;N) &=& -\bar q^{{3\over 2}\varepsilon (x_{12})}\
S^{\alpha_1\alpha_2}(x_1,x_2;N)\ \ .
\end{eqnarray}
(In deriving (2.25) we have used the properties (2.5) of $a^\pm$ which imply,
in particular,
$(a^-_{\alpha_1}a^+_{\alpha_2}-qa^+_{\alpha_1}a^-_{\alpha_2})\
{\Pi_3}^{\alpha_1\alpha_2}_{\beta_1\beta_2}=0$.)

The term ``CVO'' for $u_\pm$ is justified by the fact that they diagonalize
the
monodromy:
\setcounter{equation}{25}
\renewcommand{\theequation}{2.\arabic{equation}}
\begin{equation}
u^\alpha_\pm (x+2\pi ,N)=e^{-2\pi iL_0}\ u^\alpha_\pm (x,N)e^{2\pi
iL_0}=q^{\mp
\left( N+{1\over 2}\right)}\ u^\alpha_\pm (x,N)\ \ .
\end{equation}
(In choosing the sign of $L_0$ in the middle expression we have taken into
account the fact that $u$ depends on $x-t$.) The true justification of the
representation (2.21) stems from the possibility to express $M$, satisfying
\begin{equation}
a^-_\sigma\ M^\sigma_\beta =a^-_\beta\ q^{N+{1\over 2}},\ \ \
a^+_\sigma\ M^\sigma_\beta =a^+_\beta\ \bar q^{N+{3\over 2}}=\bar q^{N+{1\over
2}} a^+_\beta
\end{equation}
and the exchange relations (1.14), (1.15), in terms of $a^\pm$:
\begin{equation}
M^\alpha_\beta =(\bar q^{{3\over 2}}-q^{{1\over 2}}){\cal E}^{\alpha\sigma}\
a^+_\sigma a^-_\beta +\bar q^{N+{3\over 2}}\delta^\alpha_\beta\ \ .
\end{equation}
Eq.(2.27) follows from the last equation (2.9). To verify the exchange
relations we work out the Gauss decomposition (1.3) for $M$ (2.28) with the
result
\begin{equation}
d=\bar q^{{1\over 2}H},\ \ \ \ b=(1-q^2)q^{{1\over 2}H}F,\ \ \ \
c=(q^2-1)E\bar q^{{1\over 2}H}\ \ .
\end{equation}
We note that due to the non--commutativity of $b$,
$c$ and $d$ the inverse of $M_{\pm}$
(1.3) are
\begin{equation}
M^{-1}_+= q^{{3\over 4}}\ \left(\matrix{ d^{-1}& -{\bar q}b\cr \cr 0&
d\cr}\right)\ ,\ \ \
M^{-1}_-=\bar q^{{3\over 4}}\ \left(\matrix{ d&0\cr \cr -qc&
d^{-1}\cr}\right)\ \
\end{equation}
(for $db=qbd\ ,\ cd=qdc $). Eq.(1.14) is translated into
\begin{equation}
(M_\pm )^\alpha_\beta\ a^\varepsilon_\gamma =(R^\pm
)^{\alpha\rho}_{\sigma\gamma}\ a^\varepsilon_\rho (M_\pm
)^\sigma_\beta \ ,
\end{equation}
which, in view of (2.29), expresses the $U_q$ transformation law of
$a^\varepsilon$ (cf. (2.12)).

We note that the $n$--point correlation function of the chiral field (2.21)
combines together all $n$--point conformal blocks (cf. [10]).

\subsection{Factorization and monodromy for the bar sector}

In the second factor, $\bar u$, of the product (1.1) the role of the $U_q$ and
$SU(2)$ indices is reversed and we can write
\begin{equation}
\bar u^\beta_\gamma (\bar x) =\bar u^+_\gamma (\bar x,\bar N)\ \bar
a^\beta_++\bar a^\beta_-\ \bar u^-_\gamma (\bar x,\bar N)\ \ .
\end{equation}
Here $\bar a_{\pm}$ are new (independent) pairs of $q$--Bose oscillators,
\begin{equation}
[a^\varepsilon_\beta ,\bar a^{\beta '}_{\varepsilon '}]=0,
\end{equation}
such that $\bar a^\beta_-\vert 0\rangle =0=\langle 0\vert\bar a^\beta_+$ and
\setcounter{equation}{0}
\renewcommand{\theequation}{2.34\alph{equation}}
\begin{eqnarray}
\bar a^\alpha_\varepsilon\ \bar a^\beta_\varepsilon &=& q^{\pm{1\over 2}}\
(\check R^\pm )^{\alpha\beta}_{\rho\sigma}\ \bar a^\rho_\varepsilon\ \bar
a^\sigma_\varepsilon ,\\
\bar a^\alpha_-\ \bar a^\beta_+ &=& q^{{1\over 2}}\
(\check R^+ )^{\alpha\beta}_{\rho\sigma}\ \bar a^\rho_+\ \bar
a^\sigma_--\bar q^{\bar N} \ {\cal E}^{\alpha\beta}
\end{eqnarray}
where we have noted that, in view of (1.8), (1.21c),
\setcounter{equation}{34}
\renewcommand{\theequation}{2.\arabic{equation}}
\begin{equation}
P {\bar R}^\pm=\check R^\pm\ \ .
\end{equation}
The symmetry of the braid matrices $\check R^\pm$ and the relation
$(-{\cal E}^{\alpha\beta})=({\cal E}_{\alpha\beta})$ (see (1.10b)) imply that
$\bar a^\beta_\pm$ {\em satisfy exactly the same exchange relations as}
$a^\pm_\beta$.
We have, in particular,
\begin{equation}
\langle 0\vert\bar a^\alpha_-\ \bar a^\beta_+\vert 0\rangle =-{\cal
E}^{\alpha\beta}\ \ ,
\end{equation}
\setcounter{equation}{36}
\renewcommand{\theequation}{2.\arabic{equation}}
\begin{equation}
\bar a^\alpha_+\ {\cal E}_{\alpha\beta}\ \bar a^\beta_- =
-[\bar N]\ \ ,\ \
\bar a^\alpha_-\ {\cal E}_{\alpha\beta}\ \bar a^\beta_+=[\bar
N+2]\ \ ,\ \
\bar a^\alpha_\pm\ {\cal E}_{\alpha\beta}\ \bar a^\beta_\pm = 0\ \ .
\end{equation}
The definition of the $\bar u$ monodromy,
\setcounter{equation}{37}
\renewcommand{\theequation}{2.\arabic{equation}}
\begin{eqnarray}
\bar M^{-1}\ \bar u(\bar x) &=&\bar u (\bar x+2\pi )=e^{2\pi i\bar L_0}\ \bar
u
(\bar x)e^{-2\pi i\bar L_0}=\nonumber\\
&=&\bar a_-\bar q^{\bar N+{1\over 2}}\ \bar u^-(\bar x,\bar N)+q^{\bar
N+{1\over 2}}\ \bar u^+(\bar x,\bar N)\bar a_+
\end{eqnarray}
yields the expression
\begin{equation}
(\bar M^{-1})^\beta_\gamma =(q^{{1\over 2}}-\bar q^{{3\over 2}} )
\bar a^\beta_+\bar a^\sigma_-{\cal E}_{\sigma\gamma}+\bar q^{\bar N+{3\over
2}}\delta^\beta_\gamma
\end{equation}
which satisfies
\begin{equation}
\bar M^{-1}\bar a_-=\bar q^{\bar N+{3\over 2}}\bar a_-=\bar a_-\bar q^{\bar
N+{1\over 2}}\ ,\ \ \bar M^{-1}\bar a_+
=q^{\bar N+{1\over 2}}\bar a_+\ .
\end{equation}
The Borel components of $\bar M^{-1}=\bar M_-\bar M^{-1}_+$ are
\begin{equation}
\bar M_-=\bar q^{{3\over 4}}\left(\matrix{ q^{{1\over 2}\bar H} &0\cr \cr
(q^2-1)q^{{1\over 2}\bar H}\bar F& \bar q^{{1\over 2}\bar
H}\cr}\right)\ ,\
\bar M^{-1}_+=\bar q^{{3\over 4}}\left(\matrix{ q^{{1\over 2}\bar H}
&-q(1-q^2)\bar E\bar q^{{1\over 2}\bar H}\cr \cr
0& \bar q^{{1\over 2}\bar H}\cr}\right)
\end{equation}
where the $\bar U_q$ generators are given by
\begin{equation}
\bar E=\bar a^{1 *}\bar a^2 q^{\bar N_1}\ ,\quad
\bar F=\bar q^{\bar N_1}\bar a^{2 *}\bar a^1\ ,\quad
q^{\pm {\bar H}}=q^{\pm (\bar N_1-\bar N_2)}\ ,
\end{equation}
while conjugation is defined by the bar counterpart of (2.15):
\begin{equation}
\bar a^1:=q^{{1\over 2}+\bar N_2}\ \bar a^2_-\ ,\quad \bar a^2:=
-{\bar q}^{{1\over 2}+\bar N_1}\ \bar a^1_-\ \ .
\end{equation}
We recover the exchange relation (1.22) by noting the identity
\begin{equation}
(\bar M_\pm )^\alpha_\beta\ \bar a^\gamma_\varepsilon =(R^\pm )^{\alpha\gamma}
_{\sigma\rho}\ \bar a^\rho_\varepsilon (\bar M_\pm )^\sigma_\beta
\ ,\quad
\varepsilon =+ , -\ \ .
\end{equation}
We shall impose again the relation
\begin{equation}
(\bar a^{\alpha}_{\pm} )^h=0\ ,\qquad h=k+2\ ,
\end{equation}
(cf. (2.18)) thus defining the (bar)
$h^2$ dimensional Fock space $\bar {\cal F}_h$.

To sum up, we expressed the monodromy of both chiral sectors in terms of the
corresponding QUEA generators. Thus the QUEA $U_q$ and $\bar U_q$ not only
express the hidden symmetry of the (extended) WZNW model, they realize (the
monodromy) part of the dynamical variables.

\section{The $2D$ theory: weak monodromy and $U_q\otimes\bar U_q$ invariance}

We now address the question raised in the introduction: how are the expected
properties of the local (observable) $2D$ field $g$ (1.1) realized in the
extended state space of the theory?

To answer this question we need to identify the physical space
${\cal H}$. Let us denote by  $\widetilde{\cal H}$
the extended (tensor product) space generated from the
vacuum by the action of $u,\bar u, M_\pm$ and $\bar M_\pm$. Concentrating on
diagonal theories (which exist for all levels $k$ and are the only ones
present
for odd $k$ -- see [17]) we consider the $2D$ field algebra
${\cal A}={\cal A}_h(g(t,x))$
generated by polynomials in the group valued field $g=u\bar u$
and in the Lie algebra valued chiral currents $j$ and $\bar j$, and set
${\cal H}'={\cal A}\vert 0\rangle$. Then the physical space is the subquotient
${\cal H}={\cal H}'/{\cal H}''$ where ${\cal H}''\subset {\cal H}'$ is
the maximal subspace orthogonal to all vectors in ${\cal H}'$.

Note that the currents and the field are related since we can rewrite
Eq.(1.19)
in terms of $g$:
\setcounter{equation}{0}
\renewcommand{\theequation}{3.\arabic{equation}}
\begin{equation}
-{i\over 2} h\left({\partial\over\partial
x}-{\partial\over\partial t}\right)\ g(t,x)=\ :j(x-t)\ g(t,x):\\ \
\ \ {\rm etc.}
\end{equation} and the same is true for the commutation relation (1.16).

The algebra ${\cal A}$ is reducible in $\widetilde{\cal H}$. One can indeed
verify using (1.14) and (1.22) that the operator matrices
$L_{\pm}$,
\begin{equation}
(L_\pm )^\alpha_\beta :=q^{\pm{3\over 2}}\ (\bar M^{-1}_\pm
M_\pm )^\alpha_\beta
\end{equation}
commute with $g(t,x)$ (and -- trivially -- with the currents) and hence with
${\cal A}$. The space ${\cal H}'$
(and hence also the physical space ${\cal H}$)
is an eigenspace of each of
their matrix elements
so that $((L_\pm)^\alpha_\beta - {\delta}^\alpha_\beta ) {\cal H}'
= 0\ .$This becomes obvious if we compute the
products (3.2),
\setcounter{equation}{0}
\renewcommand{\theequation}{3.3\alph{equation}}
\begin{equation}
L_+=\left(\matrix{ D&(1-q^2)B
\cr \cr 0&D^{-1}
\cr}\right) ,\ \
L_-=\left(\matrix{ D^{-1}
&0\cr \cr
(q^2-1)C &D
\cr}\right)\ ,
\end{equation}
with
$$D=q^{{1\over 2}({\bar H}-H)}\ ,\nonumber $$
\begin{equation}
B= q^{{1\over 2}(H+\bar H)}F-\bar E\ q^{{1\over 2}(H-{\bar
H}+2)}\ ,
\end{equation}
$$C=Eq^{{1\over 2}(\bar H -H)}-\ q^{{1\over 2}(H+{\bar H}-2)}\bar F
\nonumber $$
and act on the vacuum vector.
Since
\begin{equation}
[C , B]=[H-\bar H]\ ,
\ \  q^{H-{\bar H}}C=q^2 Cq^{H-{\bar H}}\ ,
\ \  q^{H-{\bar H}}B=q^{-2}Bq^{H-{\bar H}}\ ,
\end{equation}
comparison between (1.3), (2.29)
and (3.3) suggests that
$CD^{-1}$, $DB$ and $q^{{\pm}(H-{\bar H})} = D^{\mp 2}\ $ should be viewed as
generators of the true $U_q$ symmetry of the $2D$ theory.

It turns out that the
gist of the matter is contained in the corresponding finite
dimensional -- oscillator algebra -- problem. We therefore proceed to describe
the subquotient ${\cal F}={\cal F}'/{\cal F}''$ where ${\cal F}'$ is the
projection of ${\cal H}'$ into the $h^4$ dimensional tensor product Fock space
${\cal F}_h\otimes\bar{\cal F}_h$ and ${\cal F}''$ is the subspace of zero
norm
vectors in ${\cal F}'$.\\ \\
{\bf Proposition 3.1}\quad {\em The bilinear form $\langle
~ , ~\rangle$
satisfying
$\langle\Phi , O\psi\rangle =\langle ^tO\Phi  , \psi\rangle$
for any $O$ in the oscillator algebra
(see (2.16), (2.17)) is
positive semidefinite on the subspace ${\cal F}'$ spanned
by vectors of the form
$$\vert n\rangle = (A^+)^n\vert 0\rangle :=
(a^+ {\bar a}_+ )^n\vert 0\rangle\  ,\ \ \  n=0,1,\dots ,2h-2\ .$$
One has
\setcounter{equation}{3}
\renewcommand{\theequation}{3.\arabic{equation}}
\begin{equation}
(N_\alpha -\bar N_\alpha ){\cal F}'=0=a^\pm\bar a_\mp{\cal F}'\ .
\end{equation}
${\cal F}'$ admits an $h$ dimensional subspace ${\cal F}''$ of zero norm
vectors. The quotient space ${\cal F}={\cal F}'/{\cal F}''$ is $h-1$
dimensional.}\\ \\
{\bf Proof}\quad To establish the second equation (3.4) we shall prove the
identity $[a^\pm\bar a_\mp,A^+ ]=0$.
Taking, to fix the ideas, the upper sign and
applying (2.34b) and (2.5), (2.9) we indeed find
\setcounter{equation}{4}
\renewcommand{\theequation}{3.\arabic{equation}}
\begin{equation}
a^+\bar a_-A^+ =
a^+_\alpha a^+_\beta ( q^{{1\over 2}}(\check R^+)^{\alpha\beta}
_{\rho\sigma}\ \bar a^\rho_+\bar a^\sigma_--\bar q^{\bar N}{\cal
E}^{\alpha\beta}) =A^+ a^+\bar a_-\ .
\end{equation}

The rest of the proof reduces to a computation of the norm square of the basis
vectors in ${\cal F}'$. To do this we first establish the commutation
relation
\begin{equation}
[A^-,A^+]=[N+\bar N+2]\ \ \ \ \ \ (A^- := a^- {\bar a}_- )\ \ .
\end{equation}
Indeed, a straightforward computation using (2.5b), (2.34b) gives
\begin{eqnarray*}
A^-A^+ &=& (q^{{1\over 2}}\ a^+_\kappa a^-_\lambda\ (\check R^+ )
^{\kappa\lambda}
_{\alpha\beta} +\bar q^N{\cal E}_{\alpha\beta})(q^{{1\over 2}}\ (\check
R ^+ )^{\alpha\beta}_{\rho\sigma}\ \bar a^\rho_+\bar a^\sigma_--\bar q^{\bar
N}{\cal E}^{\alpha\beta})=\\
&=& A^+A^-+ q^2 (\ \bar q^{\bar N}[N]+\bar q^N[\bar N]
+(q-\bar q)[N][\bar N]\  )+[2]\bar
q^{N+\bar N}
\end{eqnarray*}
which yields (3.6). As a simple corollary we derive
\begin{equation}
A^-\vert n\rangle =[n][n+1]\vert n\rangle
\end{equation}
and hence
\begin{equation}
\langle n , n\rangle =[n+1]!\ [n]!\quad
\left(\ [n+1]!=[n+1][n]!\ ,\ \ [0]!=1\ \right)\ .
\end{equation}
It follows that this norm square is non--zero for $n\leq h-2$ only.
Proposition
3.1 follows.

We are now ready to answer the first question stated in the beginning.\\ \\
{\bf Proposition 3.2}\quad {\em The field $g$ (1.1) is single valued
(monodromy
invariant) on the physical subquotient ${\cal H}$}:
\begin{equation}
( g(t,x+2\pi )-g(t,x) ) {\cal H}=0\ \ .
\end{equation}

\noindent{\bf Proof}\quad Using the factorized expressions for $u$ and $\bar
u$
we can restate (3.9) as a finite dimensional equation
\begin{equation}
(a^\varepsilon\ M\bar M^{-1}\ \bar a_{\varepsilon '}-a^\varepsilon\bar
a_{\varepsilon '}){\cal F}=0\ .
\end{equation}
Due to (2.27) and (2.40) we have
\begin{equation}
a^\pm M\bar M^{-1}\bar a_\pm =a^\pm\bar a_\pm q^{\pm (\bar N-N)},\ \
a^\pm M\bar M^{-1}\bar a_\mp =a^\pm\bar a_\mp q^{\mp (N+\bar N+2)}
\end{equation}
which allows to prove (3.10) in view of (3.4).

We come finally to the meaning of $SL_q(2)$ symmetry of the $2D$ theory.

Quantum group invariance of vacuum expectation values of products of
$g(t_i,x_i)$ follows from the observation that the $2n$--point correlation
function of $g$ is given by a sum of products of manifestly $SL_q(2)$
invariant
conformal current algebra blocks with matrix elements of powers of $A^+$ and
$A^-$. The computation of the latter (which could be the only source of
breaking the $SL_q(2)\otimes \overline{SL}_q(2)$ symmetry) only relies -- as
we
saw -- on the exchange relations (2.5) and (2.34) and on the commutativity
between $a^\varepsilon$ and $\bar a_{\varepsilon '}$, all of which are quantum
group invariant.

To sum up: the methods of covariant (indefinite metric space) formulation of
quantum gauge field theory apply to the (left--right) monodromy extended
$SU(2)$ WZNW model. They provide an understanding of monodromy and quantum
group invariance in a weak sense -- as equations valid when applied to the
physical subquotient.
Extension of these results to the $SU(n)$ models, for which
the $R$--matrices are known explicitly, appears to be straightforward [9].
Incorporation of nondiagonal models in such a canonical approach is still a
challenge.

\section*{Acknowledgements}

It is a pleasure to thank A. Alekseev, P. Kulish and especially
K. Gaw\c{e}dzki for
stimulating and enlightening discussions and Ya. Stanev for a critical reading
of an earlier version of the manuscript. LKH thanks the High Energy Section of
ICTP and INFN, Sezione di Trieste, Italy for hospitality during the course of
this work. ITT thanks the Department of Mathematics at MIT for hospitality and
acknowledges the support of a Fulbright grant 19684. This work has been
supported in part by the Bulgarian National Foundation for Scientific Research
under contract F--404.

\end{document}